\documentclass[
 reprint,
 superscriptaddress,
nofootinbib,
 amsmath,amssymb,
 aps,
 prl,
 longbibliography,
floatfix,
]{revtex4-2}

\usepackage{graphicx}% Include figure files
\usepackage{dcolumn}% Align table columns on decimal point
\usepackage{bm}% bold math
\usepackage{mathrsfs}
\usepackage{upgreek}
\usepackage{natbib}
\usepackage{indentfirst}
\usepackage{hyperref}% add hypertext capabilitie
\hypersetup{colorlinks=true,linkcolor=blue,citecolor=blue,filecolor=blue,urlcolor=blue}

\begin{document}

\title{Electric Current Control of Helimagnetic Chirality from a Multidomain State \\in the Helimagnet MnAu$_2$}

\author{Yuta~Kimoto}
\affiliation{Institute for Materials Research, Tohoku University, Sendai 980-8577, Japan}
\author{Hidetoshi~Masuda}
\email{hidetoshi.masuda.c8@tohoku.ac.jp}
\affiliation{Institute for Materials Research, Tohoku University, Sendai 980-8577, Japan}
\author{Jun-ichiro~Ohe}
\affiliation{Department of Physics, Toho University, Funabashi 274-8510, Japan}
\author{Shoya~Sakamoto}
\affiliation{Institute for Materials Research, Tohoku University, Sendai 980-8577, Japan}
\affiliation{Center for Science and Innovation in Spintronics (CSIS), Core Research Cluster, Tohoku University, Sendai 980-8577, Japan}
\author{Takeshi~Seki}
\affiliation{Institute for Materials Research, Tohoku University, Sendai 980-8577, Japan}
\affiliation{Center for Science and Innovation in Spintronics (CSIS), Core Research Cluster, Tohoku University, Sendai 980-8577, Japan}
\affiliation{International Center for Synchrotron Radiation Innovation Smart, Tohoku University, Sendai 980-8577, Japan}
\author{Yoshinori~Onose}
\affiliation{Institute for Materials Research, Tohoku University, Sendai 980-8577, Japan}
\affiliation{Center for Science and Innovation in Spintronics (CSIS), Core Research Cluster, Tohoku University, Sendai 980-8577, Japan}

\begin{abstract}
In this paper, we study the domain-wall dynamics under electric current in the helimagnet MnAu$_2$. We have found that the threshold electric current of the transition from a multidomain state to a single-chiral domain state in a magnetic field is much lower than that of chirality reversal from a single-chiral domain within certain ranges of temperature and magnetic field. The chirality after the transition depends on whether the magnetic field and electric current were parallel or antiparallel. Numerical calculations based on the Landau-Lifshitz-Gilbert equation reproduced the experimental observations. These results indicate that the domain walls are highly mobile in the helimagnet.  
\end{abstract}

\maketitle
Efficient control of magnetization is essential in spintronics, as magnetic storage and memory devices utilize two degenerate states with positive and negative magnetization in ferromagnets \cite{maekawa_concepts,chappert_emergence_spintronincs,bhatti_MRAMreview,fert_electrical_control}. One way to reverse the magnetization is through the motion of domain walls \cite{parkin_Racetrack}, which are the boundaries between two magnetic states. In this context, understanding the dynamics of magnetic domain walls has been recognized as a key issue. Thus, the current-induced dynamics of ferromagnetic domain walls have been extensively studied \cite{tatara_DW_theory_review,kumar_DWmemory_review}. Recently, antiferromagnetic materials have attracted increasing interest for magnetic memory applications \cite{jungwirth_AFMspintronics,baltz_AFMspintronics_review,han_coherentAFMspintronics,rimmler_non-collinear}, given their lack of stray fields and inherently fast spin dynamics. Therefore, antiferromagnetic domain walls have also attracted considerable attention \cite{wadley_currentDWmotion,hedrich_nanoscale_AFMDW,reimers_Mn2Au,li_chiralDW_Mn3Sn,reichlova_imagingMn3Sn,sugimoto_Mn3SnDWmotion,wu_Mn3SnDW}. 

\begin{figure}[t]
\includegraphics{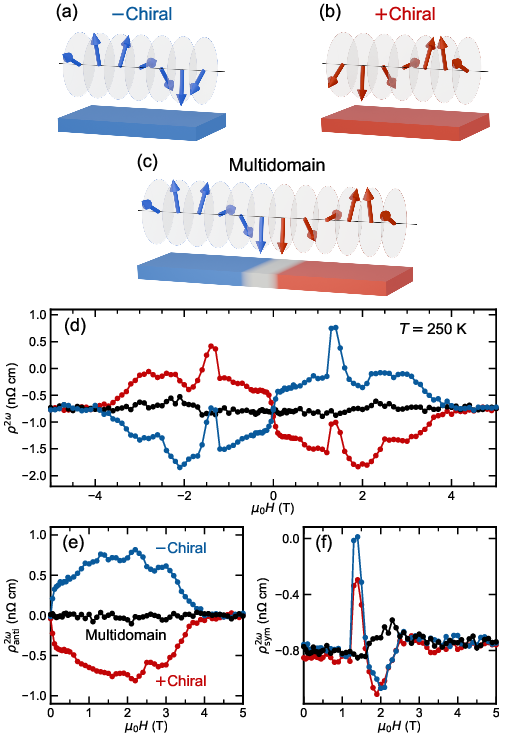}% Here is how to import EPS art
\caption{Schematic illustrations of helimagnetic single-domain states with (a) $-$Chirality and (b) $+$Chirality. (c) Schematic illustration of a helimagnetic multidomain state with a domain wall. (d) Second-harmonic resistivity $\rho^{2\omega}$ as a function of magnetic field at 250 K. Before the measurements, $+$Chiral, $-$Chiral, and multidomain states were prepared by the magnetic field sweep from $+$5 T to 0 T with applying the electric currents $+8.0\times 10^9$ A/m$^2$, $-8.0\times 10^9$ A/m$^2$, and 0 A/m$^2$, respectively. The data in the positive magnetic field region were obtained by increasing the magnetic field from the controlled initial states at 0 T, whereas the data in the negative magnetic field region were obtained by decreasing the magnetic field from the separately prepared initial states. (e) The field-antisymmetric component of second-harmonic resistivity $\rho^{2\omega}_\mathrm{anti}=(\rho^{2\omega}(+H)-\rho^{2\omega}(-H))/2$. (f) The field-symmetric component of second-harmonic resistivity $\rho^{2\omega}_\mathrm{sym}=(\rho^{2\omega}(+H)+\rho^{2\omega}(-H))/2$.}
\label{fig:FIG1}
\end{figure}
Helimagnets are a class of antiferromagnets in which the magnetic moments form a helical spin order \cite{yoshimori_MnO2}. When the crystal structure is centrosymmetric, two helimagnetic states with opposite handedness, referred to as $-$Chiral state [Fig. 1(a)] and $+$Chiral state [Fig. 1(b)], are degenerate. In this case, two chiral domains may coexist, forming domain walls \cite{hubert_DW, melnichuk_hubertWall}, which are the focus of the present study. According to previous theoretical studies, the energetically favorable domain wall is oriented perpendicular to the helical propagation vector \cite{li_vortexDW}, as shown in Fig. 1(c). Recently, it was demonstrated that the helimagnetic chirality can be controlled by decreasing the magnetic field from an induced ferromagnetic state under an electric current \cite{jiang_chiralitycontrol,masuda_chirality_control_MnAu2}. In that situation, a specific chiral domain is predominantly generated during the transition from the induced ferromagnetic state to the helimagnetic state. Nevertheless, to our knowledge, no experimental signatures of chiral domain‑wall dynamics in the multi‑chiral‑domain state have been reported so far, whereas domain walls associated with changes in the propagation‑vector direction have been reported in the literature \cite{schoenherr_topologicalDW}. 

In this paper, we report that, within certain ranges of temperature and magnetic field in the deep helimagnetic state where the full chirality reversal is impossible with accessible current magnitude, the mixed state of two chiral domains (multidomain state) can be easily converted into either a left- or right-handed single-chiral domain state by an electric current. In a multidomain state, a change in the chirality volume can be achieved only through domain-wall motion whereas reversal to the opposite chiral state requires a chirality-flip process. Therefore, we attribute this phenomenon to the mobile nature of magnetic domain walls in the helimagnet MnAu$_2$.

MnAu$_2$ is a monoaxial helimagnet with a centrosymmetric tetragonal crystal structure \cite{hall_MnAu2structure}. The helical propagation vector is $\mathbf{q}=(0, 0, 0.28)$ in the reciprocal lattice unit \cite{herpin_MnAu2neutron}. We fabricated a single-crystalline MnAu$_2$ (100 nm) thin film with a helimagnetic transition temperature $T_\mathrm{c}=340$ K on a ScMgAlO$_4$ (10$\bar{1}$0) substrate by sputtering. The $\mathrm{Mn}\mathrm{Au}_2$ thin film was grown along the [110] direction at 400 $\mathrm{^\circ C}$ and capped with a Ta (10 nm) layer, followed by 1-h post-annealing at 600 $\mathrm{^\circ C}$. The film was processed into Hall-bar devices with a channel length of 25 $\upmu\mathrm{m}$ and a width of 10 $\upmu\mathrm{m}$, where the electric current flows along the $\mathrm{Mn}\mathrm{Au}_2$ [001] direction, parallel to the helical propagation vector. Details of the sample fabrication are described elsewhere \cite{masuda_chirality_control_MnAu2}.

The helimagnetic chirality can be probed through the nonreciprocal electronic transport under a magnetic field along the helical axis \cite{jiang_chiralitycontrol,masuda_chirality_control_MnAu2}, which corresponds to the field-antisymmetric component of the second harmonic resistivity. Figure 1(d) shows the second harmonic resistivity $\rho^{2\omega}$ for $+$Chiral, $-$Chiral, and multidomain states, which were obtained by sweeping the magnetic field along the helical axis from $+$5 T (forced-FM phase) to 0 T, while simultaneously applying dc electric currents of $+8.0\times 10^9$ A/m$^2$, $-8.0\times 10^9$ A/m$^2$, and 0 A/m$^2$, respectively. For the measurement of $\rho^{2\omega}$, we utilized an ac current with an amplitude of $4.2\times 10^9$ A/m$^2$ and a frequency of 79.19 Hz induced by an electric current source (Keithley, 6221) and measured the second harmonic voltage with a lock-in amplifier (NF Corporation, LI5650). The temperature increase due to Joule heating during the measurement at 250 K was estimated to be 1.5 K from the increase in linear resistivity. While $\rho^{2\omega}$ did not exhibit discernible magnetic field dependence for the multidomain state, antisymmetric magnetic field dependences with opposite signs were observed for $\pm$Chiral states. In addition, positive peaks were observed around $H=\pm1.4$ T for both the chiral states. We decomposed $\rho^{2\omega}$ into the field-antisymmetric component [Fig. 1(e)] and the field-symmetric component [Fig. 1(f)]. The antisymmetric component clearly reflects the chirality, as reported previously \cite{masuda_chirality_control_MnAu2}. The correspondence between chirality and the antisymmetric component of second harmonic resistivity has also been confirmed for two other helimagnets, MnP and YMn$_6$Sn$_6$ \cite{jiang_chiralitycontrol,masuda_critically_YMS}. In particular, for YMn$_6$Sn$_6$, we recently succeeded in observing controlled chirality by neutron diffraction as well, which clearly indicates the correspondence between chirality and second harmonic resistivity \cite{masuda_direct_YMS}. On the other hand, a positive sharp peak around 1.4 T and a broad negative peak around 2 T were observed in the symmetric component for both $\pm$Chiral states while the magnetic field dependence was almost absent for the multidomain state. These features appear to originate from the sliding motion of helimagnetic structure \cite{kimoto_sliding}, because a kink associated with the sliding motion was correspondingly observed in the differential resistivity between 1.5 T and 2.25 T, as shown in the Supplemental Material \cite{supp}. Thus, $\rho^{2\omega}$ consists of three contributions: a field-antisymmetric nonreciprocal component, a field-symmetric component arising from sliding, and a field-independent component of $\approx -0.8$ n$\Omega$ cm. The last component originates from extrinsic effects, such as imperfect electrical contacts and/or some thermoelectric voltage originating from Joule heating. A similar contribution was observed in other samples and analyzed in detail for MnP in a previous study \cite{jiang_chiralitycontrol}. We found that this contribution depends strongly on the measurement frequency but remains constant with respect to the magnetic field (see the Supplemental Material \cite{supp}). We infer that the same holds for the MnAu$_2$ thin films. This component does not hinder the chirality detection, because the field-antisymmetric component remains nearly absent in the multidomain state, and the field-symmetric component remains nearly constant below 1.0 T and above 2.5 T. We utilize $\rho^{2\omega}$ as a probe of the chirality, given that the constant background and the effect of sliding also affects $\rho^{2\omega}$.

Then, we examine the influence of electric current application on the multidomain and $\pm$Chiral states. For this purpose, we repeatedly performed a sequence consisting of an electric current pulse application with a duration of 1 s followed by temperature stabilization and a subsequent $\rho^{2\omega}$ measurement, while varying the pulsed-current value. We also use the same current source to generate the current pulses as in the second-harmonic resistance measurements. In Fig. 2(a), the pulsed electric current was initially decreased from 0 A/m$^2$ to $-22\times 10^9$ A/m$^2$ at 0.1 T. It was then increased to $+22\times 10^9$ A/m$^2$, and subsequently decreased back to $-22\times 10^9$ A/m$^2$. While $\rho^{2\omega}$ for $\pm$Chiral states remains almost unchanged, $\rho^{2\omega}$ for the multidomain state exhibits a sudden increase as the electric current is initially decreased from 0 A/m$^2$ to $-16\times 10^9$ A/m$^2$, coinciding with $\rho^{2\omega}$ of $-$Chiral state. This indicates that the application of electric current induces a transition to $-$Chiral state. Figure 2(b) shows the result of a similar experiment with the reversed electric current at $+$0.1 T, and Figs. 2(c) and 2(d) show the experimental results under a negative magnetic field of $-$0.1 T with the original and reversed electric currents, respectively. Similar to Fig. 2(a), $\rho^{2\omega}$ for the multidomain state exhibits a sudden change around $\pm16\times 10^9$ A/m$^2$ in Figs. 2(b)-(d), while the sign of the change depends on the directions of the current and magnetic field. In Fig. 2(d), $\rho^{2\omega}$ for the multidomain state becomes comparable to that of $-$Chiral state after the sudden change, while it becomes comparable to $+$Chiral state in Figs. 2(b) and 2(c). These results indicate that the electric current induces the transition from the multidomain state to a uniform chiral state. The selected chirality depends on whether the current and magnetic field are parallel or antiparallel, which is similar to the previously reported chirality control experiments \cite{jiang_chiralitycontrol, masuda_chirality_control_MnAu2}. What we found here is that the transition from the multidomain state to a uniform chiral state can be induced even deep within the helimagnetic phase by an accessible magnitude of electric current. Since the difference in resistivity between the uniform chiral state and the multidomain state is only 0.15\% at most, the sudden change observed exclusively in the multidomain state cannot be explained by the effect of Joule heating. Instead, it should be attributed to the effect of domain-wall motion.
\begin{figure}[t]
\includegraphics{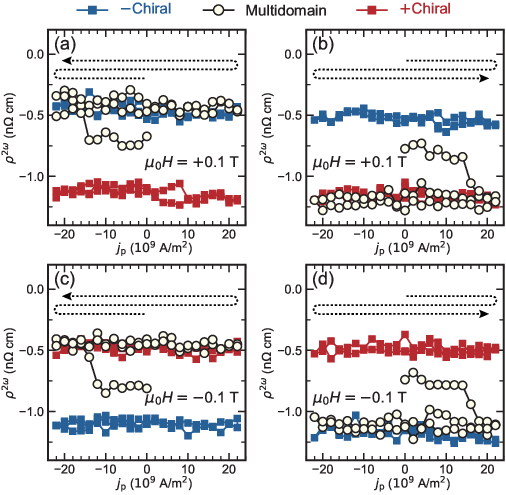}% Here is how to import EPS art
\caption{Second-harmonic resistivity $\rho^{2\omega}$ measured after the application of an electric current pulse $j_\mathrm{p}$ with a duration of 1 s at 250 K. The magnetic field was $+$0.1 T in panels (a) and (b), and $-$0.1 T in panels (c) and (d). The initial $+$Chiral, $-$Chiral, and multidomain states were prepared similarly to the case of Fig. 1. The electric current pulses were repeatedly applied while changing their magnitude and polarity, as indicated by dotted arrows, and $\rho^{2\omega}$ was measured after each pulse application. The electric current was first decreased from 0 A/m$^2$ in panels (a) and panels (c), and increased from 0 A/m$^2$ in panels (b) and panels (d).}
\label{fig:FIG2}
\end{figure}

\begin{figure*}[t]
\includegraphics{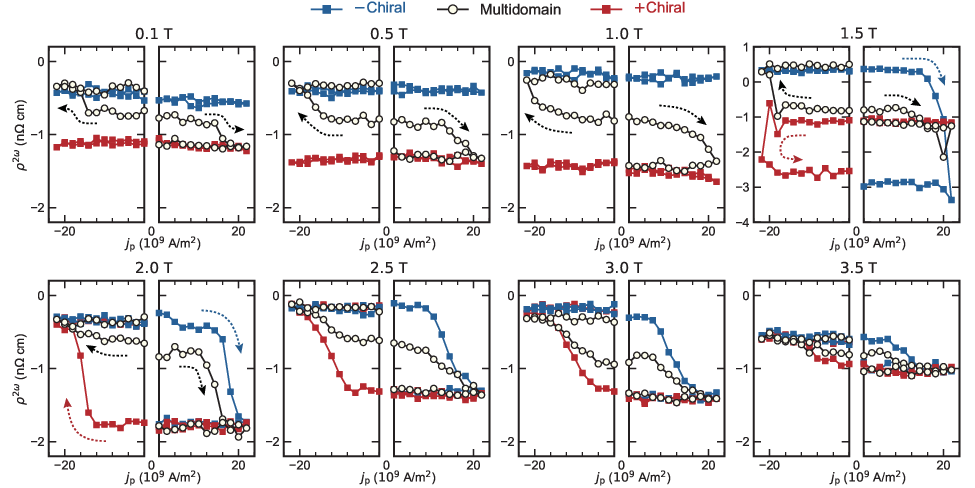}% Here is how to import EPS art
\caption{Second-harmonic resistivity $\rho^{2\omega}$ measured after the application of an electric current pulse $j_\mathrm{p}$ at various magnetic fields at 250 K. The electric current pulses were repeatedly applied while changing the magnitude, as indicated by dotted arrows. For each field, two subfigures are shown in this figure: In the left subfigure, the electric current was decreased from 0 A/m$^2$ to $-22\times 10^9$ A/m$^2$ and then increased back to 0 A/m$^2$; in the right subfigure, it was increased from 0 A/m$^2$ to $+22\times 10^9$ A/m$^2$ and subsequently decreased back to 0 A/m$^2$. The initial $+$Chiral, $-$Chiral, and multidomain states were prepared similarly to the case of Figs. 1 and 2. Note that the range of the vertical axis is larger at 1.5 T.}
\label{fig:FIG3}
\end{figure*}
Figure 3 shows the results of similar experiments at various magnetic fields. For each field, we present the results of two series of experiments involving different sequences of electric current variation. In one experiment, the electric current was decreased from 0 A/m$^2$ to $-22\times 10^9$ A/m$^2$ and then increased back to 0 A/m$^2$. In the other, it was increased from 0 A/m$^2$ to $+22\times 10^9$ A/m$^2$ and subsequently decreased back to 0 A/m$^2$. Steep changes in $\rho^{2\omega}$ for the multidomain state were also observed in the high current region at 0.5 T and 1.0 T, similar to 0.1 T, indicating that the current-induced transition from the multidomain to the uniform chiral state commonly occurs within this magnetic field range. On the other hand, $\rho^{2\omega}$ for the uniform chiral states also exhibits an abrupt change above 2.0 T. In addition to that observed for the multidomain state, $\rho^{2\omega}$ for the $+$Chiral ($-$Chiral) state shows an abrupt increase (decrease) as the electric current is decreased (increased). When the direction of current variation is reversed, the abrupt change does not occur. This behavior is consistent with the previously reported chirality reversal induced by an electric current pulse \cite{masuda_chirality_control_MnAu2}. At 1.5 T, the variation of $\rho^{2\omega}$ with electric current is complex due to the influence of sliding motion. While $\rho^{2\omega}$ for the multidomain state steeply increases (decreases) at $-20\times 10^9$ A/m$^2$ ($+18\times 10^9$ A/m$^2$), becoming comparable to that for the $-$Chiral ($+$Chiral) state as in the low-field region, $\rho^{2\omega}$ for $+$Chiral and $-$Chiral states both exhibit an abrupt decrease around $-22\times 10^9$ A/m$^2$ and $+22\times 10^9$ A/m$^2$, respectively. The resulting $\rho^{2\omega}$ values differ markedly from those of either initial state. This is likely caused by the hysteretic nature associated with the sliding motion \cite{kimoto_sliding}.

The threshold current for the chirality reversal in the high field region decreases with the magnetic field, which may be explained by the effect of Joule heating. If the electric current pulse heats the sample above the transition temperature, the chirality is expected to be selected upon cooling through the phase transition, depending on whether the current and field are parallel or antiparallel. In that case, the threshold current should decrease with the magnetic field because a magnetic field lowers the helimagnetic transition temperature. On the other hand, the magnetic field dependence of the transition from the multidomain to a single-chiral domain cannot be produced by the artifact of Joule heating. As the magnetic field is increased, a higher current is required to complete the transition. In addition, the transition from a multidomain state in the low-field region was also observed at temperatures lower than 250 K, but was absent at 300 K, as shown in the Supplemental Material \cite{supp}. The temperature dependence also contradicts the Joule heating origin. 

To theoretically investigate the electric current-driven transition from a multidomain state to a uniform chiral state, we performed numerical simulations based on a two-dimensional $1000\times10$ spin lattice using the classical $J_1-J_2$ Heisenberg Hamiltonian as
\begin{equation}
    \mathscr{H}=-\sum_\mathrm{nn}J_1\vec{S_i}\cdot\vec{S_j}-\sum_\mathrm{nnn}J_2\vec{S_i}\cdot\vec{S_j}+\sum_i\left(-\vec{H}\cdot\vec{S_i}+KS_{ix}^2\right),
\end{equation}
where the first term represents the nearest-neighbor ($\mathrm{nn}$) exchange interaction, and the second term represents the next-nearest-neighbor ($\mathrm{nnn}$) exchange interaction considered along the helical axis ($x$ axis). $\vec{H}$ is the external magnetic field, and $K$ is the single-ion anisotropy that favors an easy plane configuration in the $y$-$z$ plane. We set $J_1=62.4$ meV, $J_2=-31.2$ meV, and $K=6.24$ meV so that the ground state is the helimagnetic state with the propagation vector along the $x$ axis at zero field. While these parameters are typical of transition-metal magnets, they are not strictly tuned for the MnAu$_2$ system. The variation of the magnetic structures under electric current was numerically calculated using the Landau-Lifshitz-Gilbert equation
\begin{equation}
\begin{aligned}
\frac{\partial \vec{S_i}}{\partial t}=&-\gamma\Bigl[\vec{S_i}\times\Bigl(\vec{H}_{{\rm eff},i}+\vec{H}_{\rm temp}\Bigr)\Bigr]\\
&-\frac{\alpha}{S}\Bigl(\vec{S_i}\times\frac{\partial \vec{S_i}}{\partial t}\Bigr)+b\Bigl(\vec{j}\cdot\nabla\Bigr)\vec{S_i},
\end{aligned}
\end{equation}
where $\vec{H}_{{\rm eff},i}=-\partial \mathscr{H}/\partial \vec{S_i}$ and $\vec{H}_{\rm temp}$ is the fluctuation field that introduces the finite temperature \cite{ohe_ChiralityControl}. $\alpha$ and $\vec{j}$ are the Gilbert damping and the unit vector proportional to the electric current, respectively. The last term represents the adiabatic spin transfer torque and $b=ja^3/2eS$, where $a= 1$ nm, $S=89\mu_B$, and $j$ is the magnitude of electric current. The details of the calculation are written in the literature \cite{ohe_ChiralityControl}. The initial multidomain state was prepared by applying a strong magnetic field as large as 10 T along the $x$ direction and slowly decreasing it to $+$0.1 T without an electric current. Figure 4(b) illustrates the spatial dependence of the calculated chirality defined as
\begin{equation}
    \lambda=\left(\vec{S}\times\frac{\partial\vec{S}}{\partial x}\right)\cdot\vec{x}
\end{equation}
at the initial state. The system is divided into domains with positive and negative chiralities. The domain wall lies perpendicular to the helimagnetic propagation vector along the $x$ axis. We then swept the electric current at a rate of $|dj/dt|=5.2\times10^{18}\ \mathrm{A/(m^2s)}$ while monitoring the evolution of the domain configuration. The electric current was first decreased from 0 A/m$^2$ to $-10\times 10^{11}$ A/m$^2$ and then increased to $+10\times 10^{11}$ A/m$^2$. Figure 4(a) shows the averaged chirality $\langle\lambda\rangle$ at 0 K, 100 K, 200 K, and 300 K as a function of electric current. The temperature dependence was not significant below 300 K. Figures 4(c)-(f) illustrate the spatial distribution of the chirality at 100 K during the current sweep. The averaged chirality is nearly zero at first but exhibits a steep increase around $-3\times 10^{11}$ A/m$^2$. As shown in Fig. 4(c), domains with positive chirality expand, resulting in an increase of the averaged chirality. In other words, the domain walls move toward the negative chiral domain, as if the electric current exerts a force on the domain walls. Below $-6\times 10^{11}$ A/m$^2$, a single-domain state with positive chirality is realized [Fig. 4(d)]. When the sweep direction is reversed, the single-domain state persists up to $+5\times 10^{11}$ A/m$^2$, above which domains with negative chirality start to appear, as shown in Fig. 4(e). As the current increases further, the negative chiral domains expand [Fig. 4(f)], and a single-domain state with negative chirality is finally realized above $+10\times 10^{11}$ A/m$^2$. Importantly, the threshold current required for the transition from the multidomain state to a uniform chiral state is smaller than that for the chirality reversal between single-domain states with opposite chiralities. This qualitatively explains why the threshold current for the transition from the multidomain state is lower. Nevertheless, it should be noted that a more accurate theoretical description is needed.  As noted above, the physical parameters we employed are typical of transition metal compounds but not finely tuned to MnAu$_2$. The dimensionality of the system is also different. In addition, it is well known that impurities have a strong influence on the dynamics of domain walls, which is not taken into account at present. These factors likely cause the difference in the threshold-current magnitude and the temperature dependence between the experiment and the calculation.
\begin{figure}[t]
\includegraphics{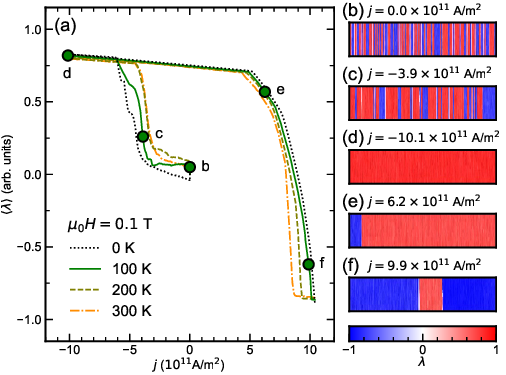}% Here is how to import EPS art
\caption{Numerical calculation of chirality variation under electric current for a $1000\times10$ two-dimensional helimagnetic model. (a) Variation of the averaged chirality under electric current at 0 K, 100 K, 200 K, and 300 K. (b)-(f) Spatial variation of the chirality at 100 K under electric currents of (b) 0 A/m$^2$, (c) $-3.9\times10^{11}$A/m$^2$, (d) $-10.1\times10^{11}$A/m$^2$, (e) $6.2\times10^{11}$A/m$^2$, and (f) $9.9\times10^{11}$A/m$^2$. The methodological details are described in the main text and ref. \cite{ohe_ChiralityControl}.}
\label{fig:FIG4}
\end{figure}

In summary, we demonstrated that an electric current with a magnitude lower than the threshold for chirality reversal drives a transition from a multidomain state, where two chiral domains coexist equally, to a uniform chiral state within a certain range of temperature and magnetic field in the helimagnet MnAu$_2$. Numerical calculations based on a two-dimensional Heisenberg model and the Landau-Lifshitz-Gilbert equation support the experimental observations. These results indicate that helimagnetic domain walls are quite mobile under electric current, implying that the pinning potential is much smaller than the energy required for domain nucleation. In addition, these observations suggest that electric current exerts an effective force on helimagnetic domain walls depending on their chirality configuration under a magnetic field, which may be useful for the manipulation of domain walls. As suggested above, the dynamics of domain walls have been extensively investigated for ferromagnets, and novel spintronic devices that exploit domain-wall motion, such as racetrack memory, have been proposed and developed \cite{parkin_Racetrack,kumar_DWmemory_review,tomasello_SkLRacetrack}. The present observation indicates that helimagnets may also exhibit similar domain-wall dynamics, potentially opening a pathway toward helimagnetic spintronics based on domain-wall manipulation. 

\begin{acknowledgments}
We appreciate T. Sasaki for her help in the film deposition. The film deposition and device fabrication were carried out at the Cooperative Research and Development Center for Advanced Materials, IMR, Tohoku University (No. 202412-CRKEQ-0402). This work was supported by JSPS KAKENHI (Grants No. JP22H05114, No. JP23H00232, No. JP23K21077, No. JP23K13654, No. JP24H01638, and No. JP24H00189), JST SPRING (Grant No. JPMJSP2114), JST PRESTO (Grant No. JPMJPR245A). Y.K. acknowledges support from GP-Spin at Tohoku University.
\end{acknowledgments}
% \begin{dataavailability}
The data are available from the authors upon reasonable request.
% \end{dataavailability}
\bibliography{Reference}% Produces the bibliography via BibTeX.
\end{document}